\documentstyle[floats,aps,epsfig]{revtex}
\begin{document}
\def \beq{\begin{equation}}
\def \eeq{\end{equation}}
\def \beqarr{\begin{eqnarray}}
\def \eeqarr{\end{eqnarray}}
\def \be{\begin{equation}}
\def \ee{\end{equation}}
\def \bea{\begin{eqnarray}}
\def \eea{\end{eqnarray}}
\def \ta{{\tilde\alpha}}
\def \tg{{\tilde g}}     
\twocolumn[\hsize\textwidth\columnwidth\hsize\csname @twocolumnfalse\endcsname
\title{Bulk and edge correlations in the compressible 
half-filled quantum Hall state}
\author{Milica V. Milovanovi\'{c}$^{1,2}$ and Efrat Shimshoni$^2$}
\address{$^1$ Department of Physics, The Technion, Haifa 32000, Israel.}
\address{$^2$ Department of Mathematics-Physics, Oranim--Haifa University,
Tivon 36006, Israel.}
\date{\today}
\maketitle
\begin{abstract}
We study bulk and edge correlations in the compressible
half--filled state \cite{HLR,rere}, 
using a modified version of the plasma analogy. The
corresponding plasma has anomalously weak screening properties, and as
a consequence we find that the correlations 
along the edge do not decay algebraically
as in the Laughlin (incompressible) case, while the bulk correlations decay
in the same way. The results suggest that due to the strong coupling
between charged modes on the edge and the neutral Fermions in the bulk,
reflected by the weak screening in the plasma analogue,
the (attractive) correlation hole is not well defined on the edge. Hence,
the system there can be modeled as a free Fermi gas of {\em electrons} 
(with an appropriate boundary condition). 
We finally comment on a possible scenario, in
which the Laughlin--like dynamical 
edge correlations may nevertheless be realized.
\end{abstract}
\vskip2pc]
\narrowtext
\section{Introduction and Principal Results}
\label{sec:intro}
Laughlin's theory of the fractional quantum Hall effect (QHE) \cite{bob} 
was given in terms of wave functions of the ground state 
and quasihole excitation. 
Using a plasma analogy to calculate the static many-body correlators, 
which characterize these wave functions, he was able to advance a very 
successful physical picture of the electron system.
The wave functions, describing the incompressible states,
contain the Laughlin-Jastrow factor, which leads to a 
special, later introduced, Girvin-MacDonald (GM) correlations in the bulk
 \cite{gmd},
and Wen's correlations on the edge \cite{wen,wentwo}. 
The Laughlin-Jastrow factor is 
everpresent in  QHE states - it exists even in the compressible 
half-filled state \cite{HLR}, for which an explicit 
wave--function has been proposed by Read and Rezayi (RR) \cite{rere}.
The question arises whether its manifestations, in terms of the above
mentioned correlations, survive in more general quantum Hall states, and in
particular in the compressible states.
Why is this question important? The correlations that are embodied in the 
Laughlin--Jastrow factor lie at the heart of various quasiparticle pictures 
\cite{cofer,cobos,dipole}
(composite fermions, composite bosons) of the QHE in the bulk. 
From the theoretical viewpoint, it is interesting to understand the status of 
Bose condensation, implicit in the Laughlin-Jastrow factor \cite{gmd,cobos},
in the compressible state. Related to this is the question 
to what extent Laughlin's quasihole construction in the compressible state 
(a zero of the wave function) can be considered as an elementary 
excitation of the system. 

Experimentally, these correlations are in principle accessible by tunneling
measurements. Indeed, recent edge--tunneling experiments by M. Grayson et al 
\cite{gray} prompted the question whether the Luttinger liquid picture 
\cite{wen,wentwo}, which is 
characterized by Wen's correlations, is valid for general quantum Hall 
systems, including the compressible states. A number of theoretical works
\cite{MZVC,leewen} have
attempted to explain the puzzling results of Ref. \cite{gray}, 
in terms of charged excitations on the edge that are effectively decoupled 
from the bulk \cite{arbnu}.  

In this paper we concentrate on the compressible QHE system at filling factor 
one-half. We assume that the RR wave function well describes the 
ground state of the system, even when we consider a system with an edge.
Namely, we assume the composite fermion (or more precisely dipole) picture 
\cite{cofer,dipole} 
to apply everywhere. We rederive the GM and  Wen's correlations  in the 
Laughlin state considering the leading order contributions of a weak-coupling 
plasma approximation (see also Ref. \cite{thesis}). 
Then we consider the same correlations (appropriately redefined) in the RR 
state. In calculating these we use the same approach -- a systematic expansion 
of a plasma free energy --  with necessary modifications to include the
Fermi sea correlations \cite{fn1}.
This introduces a statistical mechanics viewpoint of the problem, 
in terms of an anomalous, {\it weakly-screening} plasma.

Applying the forementioned procedure (and viewpoint) on the RR state, 
we find that Wen's correlations of the edge do not decay algebraically 
(at large distances) as in the Laughlin state. This excludes the possibility 
of existance of a subspace of charge density waves on the edge (of the type 
found in the Laughlin state), that is decoupled from the rest of the 
excitations -- i.e., the neutral bulk excitations. The form of the obtained 
equal-time electron Green's function on the edge suggests that, in the first 
approximation, the physical picture of the RR edge is that of a Fermi gas of 
electrons. The bulk GM correlations, on the other hand, decay 
algebraically, in an almost identical way as in the Laughlin state.

Below we detail the derivation of the correlators, in the bulk (Sec. 
\ref{sec:cb}) and on the edge (Sec. \ref{sec:ce}). 
A discussion of theoretical and experimental implications of the results is 
given in Sec. \ref{sec:summary}.

\section{Correlations of the bulk}
\label{sec:cb}
In this section, we employ the plasma analogy to derive the appropriately
generalized GM correlator in the compressible RR state. To introduce the method, 
we first use it to derive the known result for the Laughlin state (Eq. (\ref{GMCOR}) below).
\subsection{Correlations of the bulk in the Laughlin state}
In the Laughlin state, corresponding to filling factors $\frac{1}{m}$ with
$m$ odd, the GM correlator \cite{gmd} is defined as the density matrix,
\begin{eqnarray}
\lefteqn{\rho(z,z^{\prime})=}  \nonumber \\
 & &\frac{N \int d^{2}z_{2} \cdots \int d^{2}z_{N} \Psi_b(z,z_{2},\ldots,z_{N})
\times
 \Psi_b(z^{\prime},z_{2},\ldots,z_{N})}{\int d^{2}z_{1} \cdots \int d^{2}z_{N}
 |\Psi|^{2}},  \nonumber \\
 & &
\label{densmat}
\end{eqnarray}
for the bosonic many-body function,
\beq
\Psi_{b} = \prod_{i<j} |z_{i} - z_{j}|^{m} \exp\{-\frac{1}{4} \sum |z_{i}|^{2}\},
\eeq
obtained from the Laughlin wave function by omitting the phases of the relative
distances between any two electrons, $(z_{i} - z_{j})$. As shown in \cite{gmd},
the asymptotic form of $\rho(z,z^{\prime})$ is
\beq
\rho(z,z^{\prime}) \sim |z - z^{\prime}|^{- \frac{m}{2}}. \label{GMCOR}
\eeq
This correlator expresses a Bose condensation, with algebraic off--diagonal
long range order, 
of composite bosons - defined as electrons with $m$ flux quanta attached.
We now derive the above form using the weak-couplig plasma analogy.

 We first rewrite the integrand as \cite{gmd}
\begin{eqnarray}
\lefteqn{\Psi_{b}(z,\ldots,z_{N}) \times \Psi_{b}(z^{\prime},\ldots,z_{N}) =} \nonumber \\
\lefteqn{\exp\{2 m \sum^{\prime}_{i<j} \ln|z_{i} - z_{j}|\} } \nonumber \\ 
&  & \times \exp\{+ m \sum_{i}^{\prime} \ln|z - z_{i}| + 
m \sum_{i}^{\prime} ln|z^{\prime} - z_{i}|\}   \nonumber \\
&  &\times \exp\{-\frac{1}{2} \sum^{\prime} |z_{i}|^{2} - \frac{1}{4} |z^{\prime}|^{2} -
 \frac{1}{4} |z|^{2}\} \nonumber \\
& & 
\end{eqnarray}
and similarly the numerator. (The prime means that i=1 is excluded from the
summations.) Using the Laughlin plasma analogy we can write $\rho(z,z^{\prime})$ as
\beq
\rho(z,z^{\prime}) = |z - z^{\prime}|^{-\frac{m}{2}} 
\frac{Z(z,z^{\prime})}{Z(z,z)} n\; , \label{comexp}
\eeq
where $ Z(z,z^{\prime})$ is a partition function of a classical 2D plasma
with inverse temperature $\beta=\frac{2}{m}$, each particle with charge $m$,
and two impurities with charge $\frac{m}{2}$ each, at the locations $z$ and $z^{\prime}$.
($Z(z,z^{\prime})$ is a partition function with one impurity of charge $m$
at an arbitrary location, because the value of the partition function does
not depend on $z$.) $n$ is the average density of particles (equal to 
$\frac{1}{2 \pi m}$ in the usual units). To calculate the ratio of the two
partition functions we may expand the exponentials in the parameter $m$,
which we will assume to be small. The expansion will generate terms that
can be described by diagrams and corresponding rules.

As usual in this kind of expansion in the statistical mechanics analogue,
the  expansion of the denominator involves only connected diagrams.
Each diagram consists of parts, hereon called disconnected parts, which connect two
impurities at $z$ and $z^{\prime}$ but are otherwise disconnected among
themselves. Then, the rules that correspond to each diagram in the expansion are as follows:

(1) Associate with each interaction line a two-momentum satisfying momentum
conservation at each internal vertex,

(2) Associate with each interaction line between particles 
$-\frac{2 \pi \beta m^{2}}{|\vec{k}|^{2}}$, with each interaction line between
a particle and an impurity $ -\frac{2 \pi \beta m \times \frac{m}{2}}{|\vec{k}|^{2}}$,
with each interaction line between impurites $ -\frac{2 \pi \beta (\frac{m}{2})^{2}}{|\vec{k}|^{2}} $, and with each internal vertex $n$,

(3)For each incoming (from $z$) (which is also outgoing to $z^{\prime}$)
momentum for each disconnected part, integrate as $\int \frac{d^{2}k}{(2 \pi)^{2}}
\exp\{i \vec{k}(\vec{r}-\vec{r}^{\prime})\}$, but for each internal momentum as
$\int \frac{d^{2}k}{(2 \pi)^{2}}$,

(4)Multiply with a symmetry factor (if any). 
The symmetry factor is an inverse of the number of ways that we can interchange
a given number of identical parts of a given diagram, and recover the same graph.

The diagrams that represent the ineraction with the background are mutually
canceled (as we checked for the first diagrams in the expansion) and we will
not consider them. In our problem the density $n$ is fixed and depends on
the small parameter $m$. In order to get the correct order of the diagram
(i.e. the power of $m$) in the expansion, we must take this into account.
The lowest order diagram has value one. The next in order are diagrams of
the form shown in Fig. 1 and are of order $m$. We can easily sum them and the result is
\begin{eqnarray}
\lefteqn{V_{eff}(|\vec{r} - \vec{r}^{\prime}|) =} \nonumber \\
& &  (\frac{m}{2})^{2} \int 
\frac{d^{2}k}{(2 \pi)^{2}} \exp\{i \vec{k} (\vec{r} - \vec{r}^{\prime})\}
\frac{-\frac{2 \pi \beta}{|\vec{k}|^{2}}}{1 + \frac{2 \pi \beta m^{2}}{|\vec{k}|^{2}}n} .
\end{eqnarray}
The sum represents an effective screened interaction between two impurities.
The infinite summation of certain type of diagrams that diverge even more
singulary as we increase the number of interaction lines, 
is a well known ansatz in the many-body theory of the Coulomb-interacting
electron gas in three dimensions. This  captures well the phenomenon of screening
that is characteristic of long-range forces. In our case the infinite summation
is even further enforced, given the fact 
that the diverging diagrams are of the same order in $m$.
\begin{figure}[htb]
\vskip0.5in
\centerline{\epsfig{file=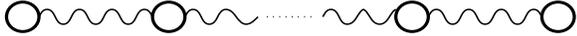,width=3in,angle=-90}}
\vskip0.5in
\caption[]
{
\label{fig:fig1}
The diagrams leading to screening of the interaction in the bulk.
}
\end{figure}

We now rewrite the ratio of partition functions in Eq. (\ref{comexp}) as
\beq
\frac{Z(z,z^{\prime})}{Z(z,z)} = \exp\{-\beta \Delta f(z,z^{\prime})\}
\; ,
\eeq
where $\Delta f(z,z^{\prime})$ represents the difference in the 
free energy between the two configurations of the impurities.
The above exponential form can be obtained by summing the set of diagrams, whose
disconnected parts are of the form shown in Fig. 1. Thus we find
\beq
\frac{Z(z,z^{\prime})}{Z(z,z)} = \exp\{V_{eff}(\vec{r} - \vec{r}^{\prime})\}.
\label{intexp}
\eeq
As $|\vec{r}-\vec{r}^{\prime}| \rightarrow \infty $ the ratio approaches unity,
because $V_{eff}$ is an effective screened interaction \cite{fn2}.
Hence, Eq. (\ref{comexp}) reduces to the well-known expression for the GM
correlator, Eq. (\ref{GMCOR}). 

This result is derived, and found to have the same form for larger, physical $m$'s \cite{gmd}.
Therefore, it is possible to analytically continue the correlator obtained
in the weak-coupling approach to larger $m$'s. Applying the same weak-coupling
infinite summation, it can be shown that the continuation is valid also in
the calculation of the static structure factor in the small-momentum limit
(when corrections to the infinite summation are added, 
this includes also the term proportional to the fourth power of the momentum 
\cite{thesis}).

It is interesting to check what does the weak-coupling approach yield for
the distribution of the charge in the tail of the Laughlin quasihole excitation
\cite{bob}. The quantity that describes this is \cite{bob}
\beq
g_{12}(|z_{1}-w|) = \frac{N}{Z} \int d^{2}z_{2} \cdots \int d^{2}z_{N}
|\prod_{i=1}^{N} (z_{i} - w) \Psi|^{2}, \label{cdenshole}
\eeq
where $\Psi$ and $Z$ is the Laughlin wave function and its norm, respectively.
In order to capture the physics of screening we sum the same, 
most imortant diagrams as before, and approximate Eq. (\ref{cdenshole})  by 
\cite{thesis}
\begin{eqnarray}
\lefteqn{g_{12}(|z_{1} - w|) = n} \nonumber \\
& & + \int \frac{d^{2}k}{(2 \pi)^{2}} 
\exp\{i \vec{k}(\vec{r}_{1} - \vec{w})\}
\frac{-\frac{2 \pi \beta (m \times 1)}{|\vec{k}|^{2}}n}
{1 + n \frac{2 \pi \beta m^{2}}{|\vec{k}|^{2}}}.
\end{eqnarray}
 As $|z_{1}-w| \rightarrow \infty$ the function
$g_{12}(|z_{1}-w|)$ should tend to the unpertubed density $n$, and it behaves as
\begin{eqnarray}
\lefteqn{g_{12}(|z_{1} - w|) = n} \nonumber \\
&  &   - m \; Const \frac{1}{\sqrt{|z_{1} - w|}} 
\exp\{- \frac{|z_{1} - w|}{r_{D}}\}\; ,
\end{eqnarray}
where $(1/r_{D}^{2})= 2 \pi \beta m^{2} n = 2 $, $r_{D}$ being the Debye length. 
\subsection{Correlations of the bulk in the compressible half-filled state}
The theory and physical picture of the filling fractions $\frac{1}{m}$ where
$m$ is {\it even}, evolved from some Fermi condensation of charged
(Chern-Simons) composite fermions (electrons with even number of flux quanta
attached) to a well-defined Fermi condensation of dipole quasiparticles
\cite{dipole}. This emphasized the advantage of Read's picture \cite{read}, which, from the begining,
takes into account the binding of electrons to (so-called) correlation hole.
(Equivalently, the statement is that the zeros of the many body functions
are found at or near the electrons). At even denominators the overall
neutral composite object is a dipole (with Fermi statistics).

The ground state wave function that corresponds to this picture is the RR wave function \cite{rere}
\beq
\Psi_{RR} = {\cal P}_{LLL} \{ \det_{i,j}[\exp(i \vec{k_{i}} \vec{R_{j}})] \Psi_{L}\},
\label{prorrwf}
\eeq
with a Slater determinant of free waves that fill a Fermi sea, 
which when projected to the lowest Landau level (LLL) 
(${\cal P}_{LLL}$ stands for the projector), acts on $\Psi_{L}$ - the Laughlin wave function.
In Eq. (\ref{prorrwf}) we wrote the determinant in terms of plane waves,
which constitute a convenient basis  for a system of free particles 
(in a rectangular geometry). The Laughlin wave function, on the other hand,
is very often expressed in the rotationaly-symmetric gauge 
(corresponding to a rotationally-symmetric geometry) amenable to the Laughlin plasma analogy.
In order to facilitate our computations we will keep these two distinct geometry
choices in the RR wave function. We justify this 
by the fact that, first, we will be interested in the (longwavelength) properties
of the system in the thermodynamic limit (when the boundary conditions should
not matter), and second, each component of $\Psi_{RR}$ will enter our
calculations in the form of translationally invariant, geometry independent elements.
 
To illustrate \cite{rere,dipole} the dipole physics contained in  Eq. (\ref{prorrwf}), 
we note that the LLL projection translates factors of the form 
$\exp\{i (k \overline{z})/2\}$, where $ k = k_{x} + i k_{y}$ and $z = x +
i y $, into the shift operator $\exp\{i k \frac{\partial}{\partial z}\}$, 
which acts on the original (before projection) holomorphic ($z$ - dependent)
part of the wave function. This effectively means that each electron becomes
displaced from the position of its correlation hole by $(-i k)$ 
(where $k$ takes values from the Fermi sea),
and therefore dipole moments are induced.

In the calculation of correlation functions the effects of the LLL projection
can be taken into account by using the following identity
\begin{eqnarray}
\lefteqn{\int d^{2}z \exp\{-\frac{1}{2}|z|^{2}\} \exp\{i \vec{q} \vec{r}\}}  \nonumber \\
& & \times \{\exp\{-i k_{1}^{*} \frac{\partial}{\partial z^{*}}\}
 \exp\{-i k_{1} \frac{z^{*}}{2}\} F_{1}(z^{*})\} \nonumber \\
& & \times \{ \exp\{i k_{2} \frac{\partial}{\partial z}\} 
 \exp\{ i k_{2}^{*} \frac{z}{2}\} F_{2}(z)\}  \nonumber \\
\lefteqn{ = \exp\{- \frac{k_{1}^{*} q}{2}\}
  \exp\{+ \frac{k_{2} q^{*}}{2}\}
  \exp\{- \frac{k_{2} k_{1}^{*}}{2}\}}   \nonumber \\
& &  \times \int d^{2}z \exp\{- \frac{1}{2}|z|^{2}\} \exp\{i \vec{q} \vec{r}\}
\exp\{-i \vec{k}_{1} \vec{r}\} \times   \nonumber \\ 
& & \exp\{i \vec{k}_{2} \vec{r}\} 
F_{1}(z^{*}) F_{2}(z)\; . 
\end{eqnarray}
If we search for the long distance behavior of the correlation functions,
usually the calculations give the same result as obtained from the 
unprojected version of the RR function.

This is the case with the appropriately generalized GM correlations to the
compressible case. The many-body wave function employed in the calculation
of the density matrix Eq. (\ref{densmat}) is
\begin{eqnarray}
\lefteqn{\Psi(z,z_{2},\ldots,z_{N}) =} \nonumber \\
& & \times \sum_{\sigma \in S_{N-1}} sgn \; \sigma
\prod_{i=2}^{N} \exp\{i (k_{\sigma (i)} z_{i})/2\} \nonumber \\
& & \times \prod_{i<j}^{\prime}  |z_{i} - z_{j} + 
i k_{\sigma (i)}-i k_{\sigma (j)}|^{2} \nonumber \\
& &  \times \prod_{i=2}^{N} |z - z_{i} - i k_{\sigma (i)}|^{2} 
\exp\{-\frac{1}{4} \sum_{i} |z_{i}|^{2}\}\; . 
\end{eqnarray}
We now introduced a particle, with coordinate $z$, without the (projected)
plane wave that enters the Fermi sea part, therefore without the fermionic
statistics that characterizes the rest of the $(N - 1)$ particles. 
The rest of its correlations with other particles is as of any other particle.
Similarly to the Laughlin case, the phase part of the Jastrow-Laughlin factor
with coordinate shifts is omitted. This, in the Chern-Simons picture, 
corresponds to attaching of two flux quanta (at distance ($i k$)) to each electron.
 
Nevertheless, as can be shown, for the type of calculations that we do,
the projection to the LLL does not affect the final result and, for the sake
of simplicity, we will explain the method on the unprojected version for which
\begin{eqnarray}
\lefteqn{\Psi(z,z_{2},\ldots,z_{N}) =} \nonumber \\
& &  \prod_{i=2}^{N} |z - z_{i}|^{2} \det \{ \exp(i \vec{k}_{i} \vec{r}_{j})
\} \prod_{i<j} |z_{i} - z_{j}|^{2} \exp\{-\frac{1}{4} \sum_{i} |z_{i}|^{2}\}. \nonumber \\
& &
\end{eqnarray}
We next assume that the dominant correlations lie in the (Jastrow-Laughlin)
differences, and for the moment neglect the Slater determinant. 
The complete plasma analogy is again possible and, as explained above, 
the infinite summation of the diagrams of type Fig.1 (for small $m$) is relevant.
In the presence of the determinant the first necessary correction to this
picture is the introduction of a new vertex, that captures also possible
Fermi (exchange) correlations between two points in the coordinate space
\cite{thesis}. In the momentum space, the vertex then corresponds to the 
static structure factor of the free Fermi gas,
\beq
s_0(\vec{q}) = n + n^{2} \int d^{2}r \exp\{i \vec{q}\cdot \vec{r}\}(g(|\vec{r}|) - 1) 
\label{ssdef}
\eeq
where $ \vec{q} \neq \vec{0} $, and the radial distribution function is
\begin{eqnarray}
\lefteqn{g(\vec{r})=} \nonumber \\
& &  \frac{1}{n^{2}} \int_{\vec{k}_{1} \in F.S.} 
\frac{d^{2}k_{1}}{(2 \pi)^{2}} \int_{\vec{k}_{2} \in F.S.} \frac{d^{2}k_{2}}{(2 \pi)^{2}}
(1 - \exp\{i (\vec{k}_{1} - \vec{k}_{2})\cdot \vec{r}\} \nonumber \\
& & \label{radfun}
\end{eqnarray}
($ F.S.$ stands for the Fermi sphere). Symbolically, the new vertex is depicted
in Fig. 2, as a sum of a direct and an exchange part, 
in which full lines represent Fermi particle lines.
\begin{figure}[htb]
\vskip0.5in
\centerline{\epsfig{file=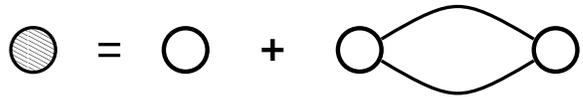,width=3in,angle=-90}}
\vskip0.5in
\caption[]
{
\label{fig:fig2}
The vertex $s_{0}(\vec{q})$ in the bulk of a RR state.
}
\end{figure}
 
From Eq. (\ref{radfun}), and the definition Eq. (\ref{ssdef}),
\beq
s_{0}(\vec{q}) - n = - \int_{R} \frac{d^{2}k}{(2 \pi)^{2}}\; .
\eeq
Here $ R $ represents the area of overlap between two Fermi spheres as shown
in Fig. 3, where the center of one of the two spheres is displaced by $\vec{q}$
from the center of the other one. The value of $ s_{0}(\vec{q}) $ 
is then given exactly by the shaded area in Fig. 3.   
The area is easily calculated for $ |\vec{q}| $ small and the result is
\beq
s_{0}(\vec{q}) = \frac{3}{4} \frac{k_{f} |\vec{q}|}{\pi^{2}}\; .
\label{smomres}
\eeq
With the necessary introduction of the new vertex $s_{0}(\vec{q})$, the interaction 
becomes less effectively screened than in the usual (Laughlin) case. It becomes
\beq
V_{eff}(|\vec{q}|) = \frac{-\frac{2 \pi \beta m^{2}}{|\vec{q}|^{2}}}
{1 + \frac{2 \pi \beta m^{2}}{|\vec{q}|^{2}} s_{0}(\vec{q})}\; ,
\label{veffmp}
\eeq
where the denominator can be interpreted as an anomalous dielectric
constant of the corresponding modified plasma.
In the coordinate space, at large distances, $V_{eff}\sim 1/r$, 
i.e. it is still long ranged and only partially screened.
\begin{figure}[htb]
\vskip0.5in
\centerline{\epsfig{file=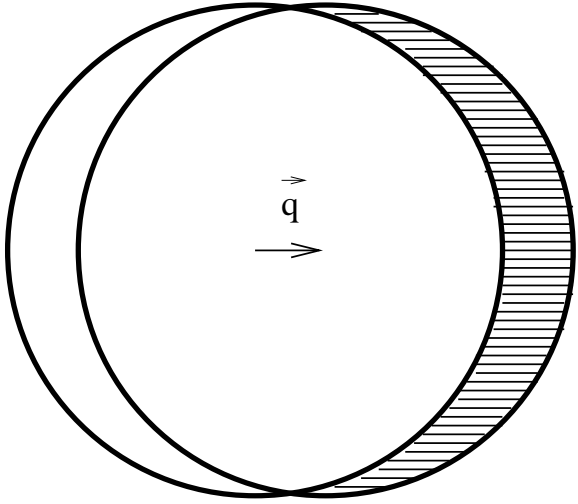,width=3in,angle=-90}}
\vskip0.5in
\caption[]
{
\label{fig:fig3}
The overlap of two shifted Fermi spheres.
}
\end{figure}

Nevertheless, if (keeping this change in mind) we apply the same 
summations and arguments as in the Laughlin case, we come up with the same
algebraic decay of the GM correlations as in that case \cite{thesis}.
This decay is slightly modified by the exponential (Eq. (\ref{intexp}))
of the partially-screened interaction (effectively a constant as in the Laughlin
case at large distances).

The question that arises immediately
is whether the analytical continuation to larger (physical) $m$'s is possible,
and, moreover, whether the screening plasma approach is reliable in giving
the leading behavior of the correlator. 
Because of the absence of a complete analogy with some physical, 
well-studied plasma, there are no available results for larger $m$ to compare with.
It was found \cite{thesis} that the weakly-screening plasma approach gives
the right (valid also for large $m$ \cite{gir}) leading (small-momentum)
behavior for the static structure factor of the compressible state, 
and generates expected (odd) powers of 
momentum in the expansion. If we try to go beyond the approach, 
and look for small- $m$ (expected) corrections, it seems that they can not
be generated \cite{thesis}. This is probably due to the nonanalyticities present in the 
compressible case (which were absent in the Laughlin case) 
that do not allow a perturbative treatment. Therefore we believe that the
approach (essentially non--perturbative) can generate the correct (large
$m$) leading behavior for the correlations that we study. They are 
between points which are directly connected only to the charge (Jastrow-Laughlin)
part of the wave function, and that immediately suggests an approach that
captures screening for their calculation.

The weak-screening property of this modified plasma can be very well seen
by considering a zero of the electron coordinates  at a point $w$ 
\cite{thesis}, which corresponds to the Laughlin quasihole
in the incompressible case,
\beq
   \prod_{i=1}^{N} (z_{i} - w) \Psi_{RR}.
\label{newqh}
\eeq
To simplify the calculation of the distribution of charge in the tail of
this excited state, we will assume the unprojected version of the RR state in Eq.
(\ref{cdenshole}) (The use of the projected state involves some complications
which are not essential and do not influence the final result). 
Now the appropriate infinite sum of the modified plasma can be expressed,
in terms of diagrams depicted in Fig. 4, with the shaded circle representing the new vertex, i.e., in this case,
\begin{eqnarray}
\lefteqn{g_{12}(|z_{1} - w|) = n} \nonumber \\
 & &  + \int \frac{d^{2}k}{(2 \pi)^{2}}
\exp\{i \vec{k} (\vec{r}_{1} - \vec{w})\}
\frac{-\frac{2 \pi \beta (m \times 1)}{|\vec{k}|^{2}}s_{0}(\vec{k})}
{1 + s_{0}(k) \frac{2 \pi \beta m^{2}}{|\vec{k}|^{2}}}\; .
\end{eqnarray}
\begin{figure}[htb]
\vskip0.5in
\centerline{\epsfig{file=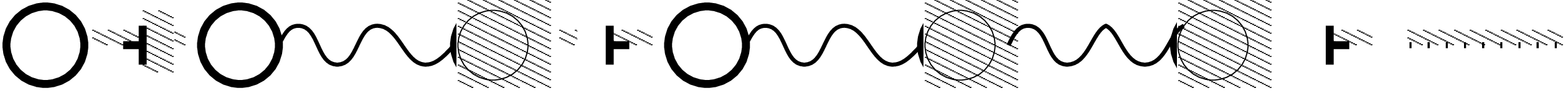,width=3in,angle=-90}}
\vskip0.5in
\caption[]
{
\label{fig:fig4}
The diagrams contributing to the calculation of $g_{12}(|z_{1} - w|)$ in the
RR state.
}
\end{figure}

The most important contributions to $g_{12}$ in the limit $|z_{1} - w| 
\rightarrow\infty$ come from nonanalyticities present in the integrand. 
They stem from the nonanalytic behavior of $s_{0}$ at $k=0$ and $k=2 k_{f}$.
Assuming that the small-momentum result Eq. (\ref{smomres}) for $s_{0}$ is
valid for any $k$ (analogously to the Thomas-Fermi approximation for the electron
gas in three dimensions) we get the contribution from the $k = 0$ region,
\beq
\{g_{12}(|z_{1} - w|) - n\}|_{T.F.} \propto - \frac{k_{f}}{m} 
\frac{1}{|z_{1} - w|^{3}}\; .  
\label{expone}
\eeq
The contribution from the $k=2 k_{f}$ region can be calculated to be
\beq
\{g_{12}(|z_{1} - w|) - n\}|_{F.O.} \propto - \frac{1}{k_{f}} 
\frac{1}{|z_{1}- w|^{3}} \sin(2 k_{f} r)\; . 
\label{exptwo}
\eeq
We may conclude from the expressions Eq.(\ref{expone}) and Eq. (\ref{exptwo}),
which summed up give the change in the distribution of the charge from the
uniform ground state contribution $n$ (in the $|z_{1} - w| \rightarrow \infty
$ limit), that the density far from the point $w$ tends to $n$ very slowly
in comparison with the Laughlin case. The charge of this excited state 
(which may be argued to be $\frac{1}{m}$ as for the Laughlin quasihole \cite{read})
is spread over a much larger region than the one in the Laughlin case, 
due to the poor-screening properties of the modified plasma.
\section{Correlations of the edge}
\label{sec:ce}
\subsection{Edge correlations in the Laughlin case}
In \cite{wentwo}, Wen showed how calculation of the equal-time correlator along the edge
in the Laughlin case can be reduced to the problem of finding the electrostatic
energy of placing an impurity outside the Laughlin plasma. 
For the sake of completeness and easy reference for our calculation we
will, in brief, repeat his arguments. We then demonstrate, that the
result is recovered in a weak coupling expansion.

{\bf Review of Wen's procedure.}
We consider a disc of the Laughlin plasma, 
with a fixed radius $R$, at fixed filling factor $\frac{1}{m}$. 
As we increase the number of particles $N$, the density will increase 
(with appropriate change in the magnetic field $B$ to keep $\frac{1}{m}$
constant) and the description that neglects detailes of the order of a magnetic
lenght would be more and more accurate, and the Laughlin plasma will behave
as a metal (with its screening properties).

To calculate the edge correlator we envision placing an impurity of charge
$m$ outside the disc of such a plasma, at a distance $z$ where $|z| \gg R$
(so that the details of the edge do not matter), and consider the ratio
$Z_I/Z$, in which
\begin{eqnarray}
\lefteqn{Z_{I}(z,\overline{z}) = \int \prod d^{2}z_{i} 
\exp\{ \sum_{i<j} 2 m \ln|z_{i} - z_{j}| \} \times} \nonumber \\
&  & \times \exp\{ + \sum_{k=1}^{N} \{- \frac{1}{2} |z_{k}|^{2} + 2 m \ln|z - z_{k}| \}\}
\end{eqnarray}
and
\beq
Z = \int \prod d^{2}z_{i} \exp\{ \sum_{i<j} 2 m \ln|z_{i} - z_{j}| + 
\sum_{k=1}^{N} \{-\frac{1}{2} |z_{k}|^{2}\}\}.
\eeq
From the first-quantization (quantum-mechanical) point of view the ratio
is the one-particle (electron) density at point $z$. On the other hand,
from the point of view of the plasma analogue, $\ln\frac{Z_{I}}{Z}$
is the electrostatic energy required to transfer the impurity 
from infinity to the point $z$. This energy can be expressed as
\beq
\ln\frac{Z_{I}}{Z} = m N 2 \ln|z| - m \ln(1 - \frac{R^{2}}{|z|^{2}}) +
o(1/N)\; .
\label{wensexp}
\eeq
The first contribution is the electrostatic energy between the total charge
$N$ and the impurity, where in the first approximation the plasma droplet is assumed
undeformed by the presence of the impurity. The second contribution describes
the most important part of the deformation that occurs: the image
charges of the impurity \cite{fn3}. The rest of the contributions are expected
to be of order $\frac{1}{N}$ or less (Due to the form of the first contributions,
analiticity in $N$ is expected.)

To find out the electron correlator between points $z_{1}$ and $z_{2}$ (on
the edge), Wen first noticed that the expression on the right-hand side of
Eq. (\ref{wensexp}) is holomorfic in $z$ and anti-holomorfic in $\overline{z}$
(outside the system), and therefore can be analytically continued, i.e.
\beq
\ln\frac{Z_{I}(z_{1},z_{2})}{Z} \approx m N \ln(z_{1} \overline{z}_{2}) -
m \ln(1 - \frac{R^{2}}{z_{1} \overline{z}_{2}})\; .
\eeq
$z_{1}$ and $z_{2}$ can be considered to be even on the edge if the final
result of the analytical continuation exists, i.e. if it is finite. 
This excludes the points $z_{1} = z_{2}$ on the edge ($|z_{1}| = |z_{2}|
= R$), where the above expression is logarithmically singular. Then, if
$z_{1} = R \exp\{i \frac{y}{R}\}$ and $ z_{2} = R $, the electron
correlator is (in the disc geometry)
\begin{eqnarray}
\lefteqn{<L|\Psi^{\dagger}(z_{1}) \Psi(z_{2})|L> \equiv} \nonumber \\
&  &  \frac{Z_{I}(z_{1},\overline{z}_{2})}{Z} \exp\{-\frac{1}{4}|z_{1}|^{2}\}
\exp\{-\frac{1}{4}|z_{2}|^{2}\}\; .
\end{eqnarray}
In the limit $\frac{y}{R}\ll 1$, where circular and rectangular
geometries are indistinguishable, this becomes
\beq
<L|\Psi^{\dagger}(z_{1}) \Psi(z_{2})|L> \sim \frac{1}{y^{m}}\; ,
\label{wenscorr}
\eeq
which coincides with the correlations on the edge obtained in the (more
familiar) bosonization approach.

{\bf Derivation of the electrostatic energy of the edge impurity in the 
Laughlin case using the weak-coupling plasma expansion.}
According to  Wen's idea, in order to find the equal--time electron correlator,
it is sufficient to compute the electrostatic energy of an 
impurity of charge $m$ at a point $z$ outside the Laughlin plasma. 
We now describe the diagramatic solution of this Statistical Mechanics problem.
To simplify the calculation, we consider a plasma which extends over the
half--plane $x\leq 0$ instead of a disc (in the thermodynamic limit,
the choice of geometry is immaterial); the impurity coordinate is $z=\xi$
(along the positive $x$--axis). The derivation of this electrostatic energy,
using the weak-coupling plasma expansion, parallels that of
the density in the bulk; i.e. calculating the electrostatic energy of a particle
interacting with a negative background - the rest of the particles \cite{fn4}. 
In the present case the system is not infinite in the $x$--direction, 
and that introduces a new type of vertex in the diagrammatic expansion.
The vertex connecting two interaction lines of momenta $q_i$, $q_f$ in the
$x$--direction, which in the infinite case is
\beq
n \delta({q}_{i} - {q}_{f})
\eeq
(where $n$ is the density), is replaced in the half-plane case by
\beq
\frac{n} {2 \pi} \{\frac{1}{-i(q_{i} - q_{f})} +
\pi \delta(q_{i} - q_{f})\}\; ,
\label{halfplanever}
\eeq
i.e. proportional to the Fourier transform of theta function,
\begin{eqnarray}
\lefteqn{\int_{-\infty}^{0} \exp\{ i(q_{i} - q_{f}) x\} dx =} \nonumber \\
&  & \int_{-\infty}^{+\infty} \theta(-x) \exp\{i(q_{i} - q_{f}) x\} dx =   \nonumber \\
&  & \frac{1}{-i(q_{i} - q_{f})} + \pi \delta(q_{i} - q_{f})\; . 
\end{eqnarray}
\begin{figure}[htb]
\vskip0.5in
\centerline{\epsfig{file=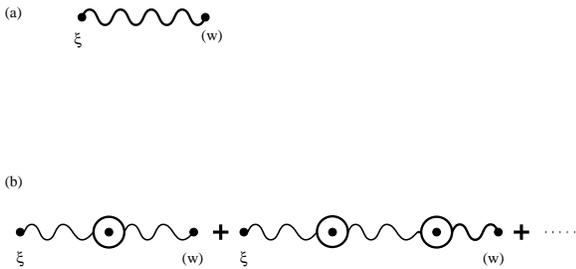,width=3in,angle=-90}}
\vskip0.5in
\caption[]
{
\label{fig:fig5}
(a) The leading contribution to the electrostatic interaction of a charged 
impurity with the plasma. (b) Diagrams that are not included in Eq. 
(\ref{wensexp}).
}
\end{figure}

The diagrams that are leading in the small $m$ expansion, and are of order
$m$, are given in Fig. 5(a) and Fig. 6. The diagram in Fig. 5(a), in which the
points $w$ in  the half--plane are integrated over, 
corresponds to the first (direct term) in  Wen's expansion, Eq. 
(\ref{wensexp}). It is also proportional to the size of the system, 
and strictly speaking diverges in the case of half-plane system. 
(This divergence does not matter, and can be handled by considering a rectangular
system with sizes $L_{x}$ and $L_{y}$ much longer than the distance ($\xi$)
of the impurity from the $y$ - axis.)

The diagrams of the type depicted in Fig. 5(b) are not included (by 
using a screened instead of the bare interaction in Fig. 5(a)),
although they are of order $m$ as well. As we remarked earlier in
this section, the diagrams that should be taken into account are of the same
form as the ones that we select to play the role of positive background
(i.e. those that cure divrgences in the expansion with 
the two-particle interaction) in the infinite system case. 
In that case, the diagram of the form in Fig. 5(a) cancels all
divergences when the interaction line does not connect to any other
interaction line. When the proper selection is done, and all diagrams that cure
divergences are present,
the complete partition function is  well-defined and a constant.
 Similary with impurities and in the semi-infinite case, 
if all due interactions (additional diagrams) are included in the partition 
function (including the interaction of impurities with positive background) 
it becomes a constant (due to the screening property of plasma). 
The partition function $Z_{I}$ in Wen's derivation is not complete, 
and therefore the part on the right-hand side of Eq. (\ref{wensexp}) is not
a constant, and can be associated with the
interaction of the impurity with ``negative background''.
\begin{figure}[htb]
\vskip0.5in
\centerline{\epsfig{file=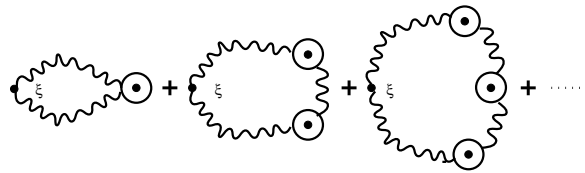,width=3in,angle=-90}}
\vskip0.5in
\caption[]
{
\label{fig:fig6}
Diagrammatic representation of the image charge energy term. The vertex 
in the half--plane case (corresponding to  Eq. (\ref{halfplanever})) is 
denoted by a dotted circle.
}
\end{figure}

The diagrams in Fig. 6 are all relevant and deserve a special attention.
Their value (at least in the long--distance limit), 
can be calculated by solving an integral equation for an effective vertex,
$V(q_{i},q_{f})$,
\begin{eqnarray}
\lefteqn{ V(q_{i},q_{f})= }  \nonumber \\
&  & \frac{n}{2 \pi} \{ [ \frac{1}{-i(q_{i} - q_{f})} + \pi \delta(q_{i} - q_{f}) ]  +   \nonumber \\
&  & \int dk [\frac{1}{-i(q_{i} - k)} + \pi \delta(q_{i} - k)] 
\frac{-4 \pi m}{(q^{2} + k^{2})} V(k,q_{f}) \}\; . 
\label{integraleq} 
\end{eqnarray}
This equation can be schematically introduced as in
Fig. 7, where we denoted only momenta along the $x$ direction. 
The momentum $q$ along the $y$ direction is the same on every line as in
the infinite-plane case. Then the contribution of all diagrams in Fig. 6,
summarized by the diagram on the left hand side of Fig. 7, can be expressed as
\begin{eqnarray}  
\lefteqn{\frac{1}{2} \int \frac{dq}{(2 \pi)} \int \frac{dq_{i}}{(2 \pi)}
\int dq_{f} \; \exp\{- i (q_{i} - q_{f})\xi \} \times } \nonumber \\
&  & \frac{- 4 \pi m}{(q^{2} + q_{i}^{2})} V(q_{i},q_{f})
\frac{-4 \pi m}{(q^{2} + q_{f}^{2})}\; . 
\label{diagtot}
\end{eqnarray}
The solution to the equation, given in the long--distance approximation,
can be found in Appendix A. It reproduces the electrostatic energy of the
imputity and its image charge in the half-plane case, corresponding to the
leading contribution to the second term in Eq. (\ref{wensexp}) when the disc
is considered to be large ($R\gg (|z|-R)$). 
\begin{figure}[htb]
\vskip0.5in
\centerline{\epsfig{file=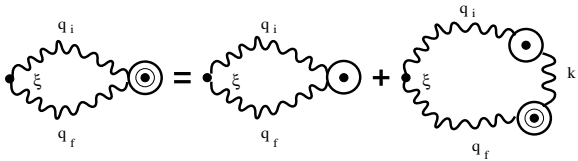,width=3in,angle=-90}}
\vskip0.5in
\caption[]
{
\label{fig:fig7}
The infinite sum of diagrams included in Fig. 6, represented as an integral 
equation for the effective vertex $V(q_{i},q_{f})$ (dotted double--circle).
}
\end{figure}

\subsection{Edge correlations in the compressible half-filled state}
Now we switch to the calculation of the edge correlations in the RR 
case using the diagrammatic method. We first consider the unprojected
RR state, and as a basis of free waves that enter the Slater determinant,
we choose
\begin{eqnarray}
&  &\frac{\exp(i k_{y} y)}{\sqrt{2 \pi}} \times \frac{\cos(k_{x} x)}{\sqrt{\pi}}
\nonumber \\
\lefteqn{{\rm  or}}  \nonumber \\
&  &\frac{\exp(i k_{y} y)}{\sqrt{2 \pi}} \times \frac{\sin(k_{x} x)}{\sqrt{\pi}}
\label{basises}
\end{eqnarray}
where $k_{x}$ and $k_{y}$ take values from a Fermi box (not sphere) in the
$k$ - space. As in the Laughlin case, we assume that the radius of the Laughlin
disc is very large in comparison with the distance (along the edge) over which
we measure correlations. So, effectively, we again consider the half--plane
problem for which, on the other hand, the basis choices (Eq.(\ref{basises})) are
also appropriate; there is no discrepancy between geometries of the Laughlin--Jastrow
and free--wave part in the ground state, as in the full-plane case. 
In Eq. (\ref{basises}) the coordinate $x$ is measured from the edge of the
half-plane, i.e. a tangent to the large disc. If, somehow, the charge and
neutral (fermionic) part decouple on the edge, the choices  
(Eq. (\ref{basises})) are quite natural, because they satisfy the requirement
that the (neutral) current normal to the boundary is zero i.e., 
that the fermionic number is conserved.

First, we consider the correlations of the object introduced in
Eq. (\ref{newqh}) to which, due to the correspondence of its construction
to the one of the Laughlin quasihole, we will refer as a quasihole. This,
of course, does not entail that the quasihole is a well-defined object --
eigenstate of the Hamiltonian, as in the case of the Laughlin quasihole.
It might be such (on the edge) if we find that its correlations are of the same
type as in the Laughlin case (Eq. 
(\ref{wenscorr})), and, therefore the charge degrees of freedom (on the
edge) in the RR state can be described in the Luttinger liquid framework
(or, microscopically, by the possible states of quasiholes). Again, as in
section     III A, to mimic the charge part of the electron, in this case, we
consider the correlations of the object constructed by putting $m \; (m =
2)$ quasiholes at the same place. Then, to find out if there is a departure
from the Laughlin case, we consider the density of this object outside the
half-plane system described by the RR state. 

In the language of the  modified plasma, we are placing a charged impurity
(not directly connected to the plane-waves part) outside the system, 
and checking whether the image charge physics still holds. Due to the poor
screening in this modified plasma, the charge induced by the external
impurity does not accumulate near the edge (within a microscopic
screening length) as in the ideal plasma. Rather, the induced charge is
expected to slowly decay towards the interior of the system. To get a
handle on the form of the electrostatic energy associated with this
effect, we can employ the Thomas-Fermi approximation to compute the induced
charge, given the dielectric properties of the modified plasma derived in
the previous section (Eq. (\ref{veffmp}) and the prodeeding
discussion). The calculation is summarized in Appendix B. We find that
unlike the Laughlin (ideal) plasma case, the leading behavior of the
``image charge" electrostatic energy is a constant, rather than a
logarithmically singular term. We now derive this result systematically
in the diagramatic expansion framework.

We first must find an effective vertex which corresponds to Eq. 
(\ref{halfplanever}) in the Laughlin case, and represented by the diagrams
in Fig. 6. That will be done at the same level of approximations as
in the case of the bulk correlations.
Explicitly, to find the value of the effective vertex, we consider the two
contributions, direct and exchange, depicted symbolically in Fig. 8, 
in the simplest diagram with only two interaction lines. (The use of 
the dotted lines is  to emphasize that we are now in the half--plane case).
\begin{figure}[htb]
\vskip0.5in
\centerline{\epsfig{file=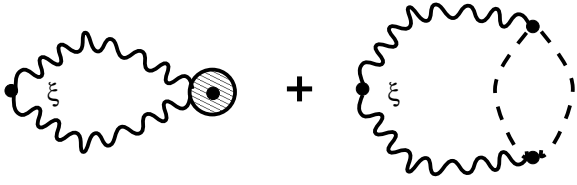,width=3in,angle=-90}}
\vskip0.5in
\caption[]
{
\label{fig:fig8}
The modified half--plane vertex in the RR state; the second diagram 
represents the exchange contribution.
}
\end{figure}

The direct contribution (unapproximated) for the first choice of basis for
the Fermi sea in Eq. (\ref{basises}) is
\begin{eqnarray}
\lefteqn{= \frac{1}{2} \int d^{2}z_{1} \ln^{2}|z - z_{1}| (2 m)^{2} \; n \;
\frac{1}{N} \sum_{\vec{k}\in F.S.} \cos^{2}(k_{x} x_{1}) = } \nonumber \\
& = \frac{n}{2} (4 \pi m)^{2} \int \frac{d^{2}p}{(2 \pi)^{2}}
\int \frac{d^{2}q}{(2 \pi)^{2}} \times & \nonumber \\
& \frac{1}{p_{x}^{2} + p_{y}^{2}} \frac{1}{q_{x}^{2} + q_{y}^{2}}
\exp\{-i (q_{x} - p_{x})z\} (2 \pi) \delta(q_{y} - p_{y}) \times & \nonumber \\
& \int dx_{1} \exp\{i (q_{x} - p_{x}) x_{1}\} \frac{1}{N} \sum_{\vec{k} \in
F.S.} \cos^{2}(k_{x} x_{1}) &
\end{eqnarray}
where the summation in $\vec{k}$ runs over the Fermi box. 
This summation can be rewritten as
\beq
\frac{1}{N} \sum_{\vec{k} \in F.S.} \cos^{2}(k_{x} x_{1}) = \frac{1}{2} +
\frac{1}{N} \sum_{\vec{k} \in F.S.} \frac{\cos(2 k_{x} x_{1})}{2}\; .
\eeq
The second part can be neglected, because it leads to an effective smearing
of both the $\delta$ -- function $\delta(p_{x} + q_{x})$, and the pole 
$\sim 1/(p_{x} + q_{x}) $ that we would get if it were a constant. 
The first part is the most important and singular 
in the infrared limit, which dominates the infinite 
summation above. Therefore, the direct contribution to the effective vertex is
\beq
\frac{1}{2} \frac{n}{2 \pi} \{ \frac{1}{- i (q_{x} - p_{x})} + \pi \delta(q_{x}
- p_{x})\}\; ,  \label{verone}
\eeq
i.e., half of the vertex in the Laughlin case.

Applying similar arguments, that is keeping the most important terms that
contribute to the value of the diagram in the long--distance limit, 
we find that the exchange contribution to the effective vertex is
\beq
(- \frac{1}{4}) \frac{n}{2 \pi} [\pi \delta(p_{x} - q_{x})]\; . 
\label{vertwo}
\eeq
We get the same contributions, Eq. (\ref{verone}) and Eq. (\ref{vertwo}),
for the second choice of basis in Eq. (\ref{basises}).
Therefore, the effective vertex in the RR case that parallels Eq. 
(\ref{halfplanever}) in the Laughlin case, is
\beq
\frac{1}{2} \frac{n}{2 \pi} \{ \frac{1}{- i (q_{x} - p_{x})} + 
\frac{\pi}{2} \delta(p_{x} - q_{x}) \}\; .
\eeq
It is not a simple multiple of the Laughlin vertex; because of the 
exchange contribution, the solution of a new integral equation, 
corresponding to Eq. (\ref{integraleq}) with the new ``bare'' vertex,  
will not yield the leading, logarithmic behavior characteristic of the Laughlin
case, which can be translated into the algebraic decay of the quasihole correlator.
Namely, if $a \neq 1$ (for the definition of $a$ see Appendix A), 
the solution of the integral equation  in the long--distance limit is
\begin{eqnarray}
\lefteqn{V(q_{i},q_{f}) \approx }   \nonumber  \\
& - \frac{(q^{2} + q_{i}^{2})}{b} \delta(q_{i} - q_{f}) + &  \nonumber \\
&  \frac{1}{-i(q_{i} - q_{f})} \frac{c}{(\pi \; a \; c \; b)^{2}} 
(q^{2} + q_{i}^{2}) (q^{2} + q_{f}^{2})\; . &
\end{eqnarray}
In our case $b = - 4 \pi m $, $a = \frac{1}{2}$, 
and $c = \frac{1}{2} \frac{n}{2 \pi}$. The contribution from 
the $\delta$-- function to the electrostatic energy is
\beq
- m \ln \frac{\Lambda}{q_{c}},
\label{leadbeh}
\eeq
where $q_{c}$ is the infrared cut-off. There is no obvious way to cancel
the $q_{c}$-dependence; i.e., in the case of the modified  plasma, 
we must keep the size of the system finite, and the distance of the impurity
should be considered smaller than the size of the
system, in the calculations. (Note that this appear to indicate, that 
the expansion in $N$ is not analytic as  in the Laughlin case.)
The contribution of the second term can be written as
\beq
+ m \; Const \times (\Lambda - q_{c}) q_{c} \times (q_{c} \xi) + f(\Lambda,q_{c}) +
o((q_{c} \xi)^{2})
\eeq
where $Const > 0$, and $f$ is an algebraic function of $\Lambda$ and $q_{c}$.
In principle, other contributions, of order higher in $m$, 
from a more detailed solution of the integral equation can be calculated.
We expect that their dependence on $\xi$ will be of the form $(\xi q_{c})^{n}$
or $(\frac{1}{\xi})^{n}$, where $n$ takes on positive integral values, 
$(\xi < \frac{1}{q_{c}})$, and will not change the leading behavior 
Eq. (\ref{leadbeh}) (in which we are interested the most).
In the scope of our approach, 
which takes $m$ small (and assumes the possibility
of an analytical continuation to higher $m$), it is hard to estimate the
true coefficients in front of the powers of $\xi$, due to the requirement to
know them to all orders in $m$. Also, there might be relevant contributions
from other diagrams (in the small $m$ expansion) which we did not consider.
But as we assume that the plasma correlations (although modified) 
are dominant for the calculation of the quasihole correlator, we do not expect
that there will be any change in the leading behavior described by Eq. (\ref{leadbeh}).

The above calculations imply that the overlap between  
two quasihole excitations
on the edge does not depend on the distance between them; it is a constant,
but decreases with the size of the system. This might be understood,
taking into account that the quasiholes in the RR state are not well--defined,
well--localized objects in the bulk  (see the end of section II B), and certainly
not on the edge where the screening seems to be even weaker than in the bulk.

Once we take this point of view that, in fact, the states described by
the Laughlin quasihole construction on the edge are extended, a special
care must be taken concerning their normalization. In general, the
normalization is expected to depend on the size of the system (as in the
case of the free waves (in the non--interacting system)). Therefore, the
first contribution to the plasma electrostatic energy 
(Eq. (\ref{leadbeh})) (that
through the infrared cut-off depends on the size of the system) might be a
consequence of an incomplete normalization of the quantum--mechanical
correlator at the beginning of our calculation. If this term is included
(in the normalization) from the beginning, the value of correlator at
large distances (in our approximation) approaches unity \cite{fnwen}.

To find out the electron correlator, we must take into account the correlations
that come from the neutral (plane-wave) part of the RR function, alone.
These are not included in the preceding (modified plasma) calculations,
which gave the correlator of the quasihole, 
the object that (in our approximation)
carries the charge part of the electron. 
The neutral contribution is expected to be of the form
\beq
\sim g \frac{\sin(k_{y}^{F} y)}{y}\; ,
\eeq
where $g$ is a coefficient that depends on the boundary conditions. 
When combined with the charge correlator, 
it produces the usual (physical) decay of the electron correlator with the
distance. Except for the dependence on the size of the system, 
the electron correlations on the edge are as if the system was a free 
(two-dimensional) Fermi gas of {\em electrons}. 
It can be shown that the same long-distance
behavior of the correlations follows from the LLL projected 
wave function (Eq. (\ref{prorrwf})).

\section{Discussion and Conclusions}
\label{sec:summary}
If we assume that, indeed, the whole description of the edge of the compressible state is equivalent to that of a free Fermi gas, we can try to predict the occupation numbers (probability density) of electron near the edge. Then the second choice for the 
boundary condition in Eq. (\ref{basises}) is more appropriate because the probability density $(\rho (x))$ should vanish at some point near the edge $( x \sim 0 )$. The resulting probability-density distribution
\beq
\rho(x) \sim [ k_{F} - \frac{\sin(2 k_{F} x)}{2 x} ] \; \; \; (x < 0),
\eeq
is very similar to the smooth function that one can get extrapolating the data that describe the occupation numbers for electron near the edge in (finite-system) exact-diagonalization studies \cite{yanghan}, and the observed oscillations might be identifi
ed as the Friedel oscillations. Also, with the above assumption,
 the  density of
states for electron tunneling into the compressible edge would be similar
to the one for tunneling into a Fermi liquid (metal). This is consistent
with our intuitive expectations given the compressible nature of the system,
if, loosely speaking, the characteristic energies for the motion of the charge
and neutral (Fermi) part are comparable.

We believe that it would be possible to construct an effective $1 + 1$ 
dimensional theory along the edge, which has the same correlations that we expect,
taking the coordinate normal to the edge to corresponds to time i.e. $x \equiv
vt$, and translating our diagramatic calculations into an effective interaction
between a neutral and charged part. This would yield a model for the 
suppression of the correlation of the chiral boson theory \cite{wen,wentwo}
(charge part), which assumes
that its neutral and charge components move with the same velocity 
$(v_{c} = v_{n} = v)$ along the edge. If the model is generalized to the
one for which $v_{c} \gg v_{n}$, at sufficiently large momenta--high
energies where the exchange part of the interaction is suppressed (due to
a reduced overlap of the two one--dimensional spheres), we expect that the chiral boson
correlations will be released. Therefore, the difference in the dynamics
of the charge and neutral part appear to be a necessary condition for
the decoupling of the edge and bulk (the charge and neutral part) 
at high enough energies, as seen in experiments \cite{gray}.
(For a similar explanation of the simultaneous suppression of the neutral
part see \cite{leewen}.) The ``true'' (low-energy) correlations should reflect the
compressible nature of the system.

In contrast with the edge problem, the bulk correlations in the compressible
case that we considered seem to be similar to the ones in the incompressible
case. The GM correlations are almost identical, and, due to the finite screening,
the quasiholes (correlation holes) have a chance to be considered as well-defined
(albeit very extended) objects (like skyrmions when the compressible degree
of freedom -- spin -- is included in the incompressible problem \cite{sondhi}). 
This, intuitively, gives an additional support to the quasiparticle pictures
of the bulk that we have by now. On the other hand, the edge correlations
differ completely from the ones in the incompressible case. 
In incompressible states the edge physics is a reflection of the bulk physics,
and the same quasiparticle picture of the bulk is possible on the edge.
In the compressible case, and, in the plasma analogy,due to the very weak screening on the edge, we probably
can not talk about existence of the correlation hole which, 
in Read's picture of the bulk, attracts an electron, and creates a weakly--interacting
composite object -- a Fermi quasiparticle. In the scope of our approach,
and in the first approximation, electrons are unbounded and the edge of the
compressible state appears to be similar to the edge of free electron gas
(with an appropriate boundary condition).

\acknowledgements

We gratefully acknowledge useful discussions with A. Auerbach, J. Feinberg,
J. H. Han, A. MacDonald, R. Rajaraman, S. Sondhi, A. Stern and X.-G. Wen, 
and especially with N. Read. 
This work was partly supported by grant no. 96--00294 
from the United States--Israel Binational Science Foundation (BSF), Jerusalem,
Israel, and the Technion -- Haifa University Collaborative Research Foundation.
M. M. also acknowledges support from
 the Fund for Promotion of Research at Technion, and the Israeli 
Academy of Sciences. E. S. would like to thanks the hospitality and support of
the Institute of Theoretical Physics in UCSB, Santa Barbara, where part of this
work has been carried out.

\appendix
\section{}
\label{app:A}
We consider the integral equation (\ref{integraleq}) for the case of a general vertex
\beq
c \{\frac{1}{-i(q_{i} - q_{f})} + a \pi \delta(q_{i} - q_{f})\}\; .
\eeq
Then, the integral equation can be rewritten as
\begin{eqnarray}
\lefteqn{V(q_{i},q_{f}) \; [1 - \frac{\pi a \; c \; b}{q^{2} + q_{i}^{2}}] = }
 \nonumber \\
&  &  = c [ \frac{1}{- i (q_{i} - q_{f})} + \pi a \; \delta(q_{i} - q_{f}) + 
 \nonumber \\
&  & b \int dk \frac{V(k,q_{f})}{i (k - q_{i}) (k^{2} + q^{2})} ] 
\end{eqnarray}
where $ b = - 4 \pi m $. If we try simply to iterate the equation in the limit
when $ q_{i} \rightarrow q_{f}$ we find that each iteration produces a solution of the form
\begin{eqnarray}
\lefteqn{V(q_{i},q_{f}) = }  \nonumber \\
&  & \frac{\alpha(q_{i},q_{f};q)}{-i(q_{i} - q_{f})} +  \nonumber \\
&  & \beta(q_{i};q) \delta(q_{i} - q_{f}) + f(q_{i},q_{f};q) 
\label{solutionform}
\end{eqnarray}
where $\alpha$ and $\beta$ are fixed, i.e. do not change after some itrations,
and $f(q_{i},q_{f};q)$, keeps changing, but does not have any (new) singularity
as $q_{i} \rightarrow q_{f}$. It can be assumed, from the iteration analysis,
that $f(q_{i},q_{f};q)$ is analytic in  all of its variables in the long distance limit.

If we assume that the solution is of the form \ref{solutionform}, 
and that $\alpha$ is an analytic function of its variables, the integration
on the right hand side can be done, and yields a new expression
\begin{eqnarray}
\lefteqn{= c \{ \frac{1}{-i(q_{i} - q_{f})} [1 + \frac{\beta \; b}{q^{2} + q_{i}^{2}}] + 
 \pi a \; \delta(q_{i} - q_{f}) +}   \nonumber \\
&  & b \; \pi [ \frac{f(q_{i},q_{f})}{q^{2}_{i} + q^{2}} + \frac{f(i q,q_{f})}{iq
\; (iq - q_{i})} ]   +  \nonumber \\
&  &  b \; \pi \frac{1}{-i(q_{i} - q_{f})} [ \frac{\alpha(q_{i})}{q_{i}^{2} +
q^{2}} - \frac{\alpha(q_{f})}{q_{f}^{2} + q^{2}} ] +   \nonumber \\
&  & b \; \pi \frac{\alpha(iq)}{q (iq - q_{i})(iq - q_{f})} \} 
\end{eqnarray}
where $\alpha(k) \equiv \alpha(k,q_{f};q)$.
In order to equate the $\delta$--functions on both sides, (at $q_{i} = q_{f}$),
$\beta$ must be
\beq
\beta = \frac{\pi \; a \; c}{[1 - \frac{\pi \; a \; c \; b}{q^{2} +
q_{i}^{2}}]}\; .
\eeq
Then we equate the coefficients with $ \frac{1}{-i (q_{i} - q_{f})} $ 
at the point $q_{i} = q_{f}$, to get
\beq
\alpha(q_{i},q_{f};q)|_{q_{i}=q_{f}} = \frac{c (q^{2} +q_{i}^{2})^{2}}
{(q^{2} + q_{i}^{2} - \pi \; a \; b \; c)^{2}}\; .
\eeq
To be consistent with the iteration result (and also with symmetry arguments),
\beq
\alpha(q_{i},q_{f};q) = c \frac{q^{2} + q_{i}^{2}}{(q^{2} + q_{i}^{2} - \pi \; a \; b \; c)} 
\frac{q^{2} + q_{f}^{2}}{(q^{2} + q_{f}^{2} - \pi \;a \; b \; c)} 
\eeq
although it is not consistent with our assumption that $\alpha$ is an 
analytic function at the begining of the substitution of Eq. (\ref{solutionform}).
Still, it does satisfy the assumption in its long distance version
\beq
\alpha(q_{i},q_{f};q) \approx \frac{c}{(\pi \; a \;b \; c)^{2}} (q^{2} +
q_{f}^{2}) (q^{2} + q_{i}^{2})\; ,
\eeq
and the same approximation must be employed in the previous equations.

To complete the solution we must find $f(q_{i},q_{f};q)$ from the remaining equation
\beq
[1 - \pi (a + 1) \; b \; c \frac{1}{q^{2} + q_{i}^{2}}] f(q_{i},q_{f}) = 
\pi \; c \; b [\frac{f(iq,q_{f})}{iq (iq - q_{i})} ]
\eeq
(We used $\alpha(iq) = 0$ which is consistent with the fact that the 
poles at $ k = \pm iq $ in $\frac{\alpha(k)}{k^{2} + q^{2}} $ at the begining
of the calculation were spurious.) In the long--distance (or small--momentum)
approximation, i.e. when
\beq
(a+1) \frac{f(q_{i},q_{f};q)}{(iq + q_{i})} \approx \frac{f(iq,q_{f};q)}{iq}
\eeq
a nontrivial (nonzero) solution exists only when $a = 1$ (Laughlin case).
It is
\beq
f(q_{i},q_{f};q) = q_{i} + i q
\eeq
in the limit when $q_{i} \rightarrow q_{f}$ (irrespective from the value of $c$).

By power counting or by explicit calculation we can find out that the 
leading contribution in the Laughlin case comes from the latter part of the solution.
An introduction of an infrared cut-off is necessary, but the dependence on it
disappears when the $\delta$ - function part of the solution is included. 
When substituted in Eq. (\ref{diagtot}), this yields
\beq
- m \ln(\xi \Lambda)
\eeq
as the electrostatic energy of the 
impurity and its image counterpart in the longdistance approximation, 
where $\Lambda$ is an ultraviolet cut-off (corresponding, e.g., to the inverse
screening length of the plasma). There is no dependence on the infrared cut-off, 
because we are considering a half--plane (semi--infinite) system, 
and are recovering the well known result for that case.

\section{``Image Charge" Interaction Energy in a Modified Plasma}
\label{app:B}
We consider a point--like impurity of charge $m$, placed at a distance $\xi$ from
the edge of a two-dimensional modified plasma which occupies the
half--plane $x\leq 0$. The plasma is characterized by a wave--vector dependent
dielectric constant of the form
\beq
\epsilon(q)=1+{q_0\over q}\; ,\quad {\rm where}\quad 
q_0=\frac{3}{4} \frac{k_{f} }{\pi^{2}}
\label{dieconst}
\eeq
and $q=|\vec{q}|$ (see Eqs. (\ref{smomres}) and  (\ref{veffmp})). 
The electrostatic potential generated by the charge distribution, 
that the external impurity induces in the plasma, is given by 
\beq
V_{ind}(\vec{r})=V_{sc}(\vec{r})-V_{ex}(\vec{r})\; ;
\eeq
here $V_{sc}(\vec{r})$ is the screened potential of the impurity
\begin{eqnarray}
V_{sc}(\vec{r})=m\int d^2 r' D(\vec{r},\vec{r'})
\ln |\vec{r}-\xi\hat{x}|\; ,\quad {\rm where} \nonumber \\
D(\vec{r},\vec{r'})=
\theta(-x)\theta(-x')\int {d^2 q\over (2\pi)^2}
e^{-i\vec{q}\cdot(\vec{r}-\vec{r'})}\epsilon^{-1}(q)\; ,
\end{eqnarray}
and $\epsilon(q)$ is given by Eq. (\ref{dieconst}) (the Theta functions
restrict the screening to the half--plane occupied by the plasma). 
This yields (in $q$--space) 
\beq
V_{ind}(\vec{q})=-{m\pi q_0e^{-|q_y|\xi}\over (q+q_0)|q_y|(|q_y|-iq_x)}\; .
\eeq
The (two--dimensional) Poisson equation then relates this component of the  potential to 
the induced charge
\beq
\rho_{ind}(\vec{q})={q^2\over 2\pi}V_{ind}(\vec{q})=-{mq_0\over 2}
{(|q_y|+iq_x)e^{-|q_y|\xi}\over (q+q_0)|q_y|}\; .
\label{indch}
\eeq
The Fourier transform of Eq. (\ref{indch}) yields
\beq
\rho_{ind}(\vec{r})=-\theta(-x){m\over 4\pi q_0}{\cal
R}(r^2-\xi^2+2i\xi y)^{-3/2}\; ,
\eeq
where ${\cal R}$ denotes the real part, and $r^2=x^2+y^2$. Note that 
this charge distribution decays algebraically towards the interior of
the plasma, indicating its anomalously poor screening properties.
The electrostatic energy associated with the interaction of the
impurity and the induced charge is then found to be (to leading order
in small $\xi$)
\beq
E_{el}\approx -{m^2\Lambda\over 4q_0}\ln \frac{\Lambda}{q_{c}}\; .
\eeq
The higher order corrections decrease as a  function of $\xi$. Multiplying
by the inverse temperature $\beta=2/m$, and with the appropriate
definition of the ultraviolet cutoff $\Lambda$, this result coincides with Eq.
(\ref{leadbeh}), and hence is consistent with our diagramatic approach.


\begin{references}
\bibitem{fn2}
This can be compared with the exact value in the case $m=1$, 
for which $ \beta \Delta f_{\infty} = - 0.03942 $ \cite{gmd}.
(The case $m=1$ is the upper limit for the expansion that we consider.)
\bibitem{fn3}
This term has the expected singularity as the edge ($|z| = R$) and reduces to 
$ - m \ln\frac{2(|z| - R)}{R} $, as the radius of the disc is increased.
\bibitem{fn4}
In that case the contributions of the diagrams multiplied by a minus sign
should sum up to account for the 
positive background potential $\sim \frac{1}{2} |z_{i}|^{2}$.
\end{references}
\end{document}